\documentclass[conference,letterpaper]{IEEEtran}

\usepackage[utf8]{inputenc} 
\usepackage[T1]{fontenc}
\usepackage{url}
\usepackage{ifthen}
\usepackage{cite}
\usepackage{amsthm} 

\usepackage{filecontents}
\usepackage[table,xcdraw]{xcolor}
\newtheorem{theorem}{\textbf{\text{Theorem}}}
\newtheorem{corollary}{Corollary}
\newtheorem{conjecture}{Conjecture}
\newtheorem{lemma}{Lemma}

\usepackage{epsfig}
\usepackage{times}
\usepackage{verbatim}
\usepackage{enumerate}
\usepackage{multicol}
\usepackage{afterpage}
\usepackage{wrapfig}
\usepackage{cite}
\usepackage{graphicx}
\usepackage[caption=false]{algorithm}
\usepackage{algorithmic}
\usepackage{ifmtarg}
\usepackage{amssymb}
\usepackage{amsmath}
\usepackage{color}
\usepackage{bbm}
\usepackage{epstopdf}
\usepackage{xfrac}
\usepackage{flushend}
\usepackage[font=small]{caption}
\usepackage{subcaption}

\newtheorem{assumption}{Assumption}
 \usepackage{multirow}
 \usepackage{stackengine}

  \newcommand*{\bH}{\boldsymbol{H}}
  \newcommand*{\bp}{\boldsymbol{p}}
    \newcommand*{\bu}{\boldsymbol{u}}
        \newcommand*{\bv}{\boldsymbol{v}}
\begin{document}
\title{Scaling Laws of Dense Multi-Antenna Cellular Networks} 


	\author{
		\IEEEauthorblockN{\large Ahmad AlAmmouri, Jeffrey G. Andrews, and Fran\c cois Baccelli}
		\IEEEauthorblockA{The University of Texas at Austin\\
			Austin, TX 78701 USA\\
			Email: \{alammouri@utexas.edu, jandrews@ece.utexas.edu, francois.baccelli@austin.utexas.edu\}}}


\maketitle

\begin{abstract}
 We study the scaling laws of the  signal-to-interference-plus-noise ratio (SINR) and the area spectral efficiency (ASE) in multi-antenna cellular networks, where the number of antennas scales with the base station (BS) spatial density $\lambda$, under the assumption of independent and identically distributed (i.i.d.) channels. We start with the MISO case with $N_t(\lambda)$ transmit antennas and a single receive antenna and prove that the average SINR scales as $\sfrac{N_t(\lambda)}{\lambda}$ and the average ASE scales as $\lambda\log\left(1+\sfrac{N_t(\lambda)}{\lambda}\right)$. For the MIMO case with single-stream eigenbeamforming and $N_r(\lambda) \leq N_t(\lambda)$ receive antennas, we prove that the scaling laws of the conditional SINR and ASE are agnostic to $N_r(\lambda)$ and scale exactly the same as the MISO case. Hence, deploying multi-antenna BSs can help maintain non-zero per-user throughput and a corresponding linear increase in the ASE in dense cellular networks.  
\end{abstract}

\section{Introduction}
  
  
Earlier academic works \cite{A_Andrews11} and several decades of industry deployments indicate that area spectral efficiency (ASE) - the network sum throughput per unit area - increases about linearly with network densification, namely the base station density,  In contrast, we recently showed that under natural assumptions on the network and the signal propagation models, the signal-to-interference-plus-noise ratio (SINR) degrades to zero and the ASE saturates to a finite constant in the limit of very dense networks \cite{A_AlAmmouri19}. Practically, this result means that densifying the network beyond a certain point actually sacrifices the per-user performance, and in fact the per-user rate drops asymptotically to zero, a result also pointed to by other recent works on ultra-dense networks \cite{Downlink_Zhang15,Performance_Ding17,Performance_Nguyen17,SINR_AlAmmouri17}.  All those works were on single antenna transmission and reception.  In this work, we study whether deploying multi-antenna BSs can improve the scaling laws of the SINR and the ASE in cellular networks, in particular when the number of antennas is scaled with the BS density. The motivation behind this idea is the hope that multi-antenna BSs can improve the per-user SINR by increasing the gain and/or mitigating the network interference.  

 The history of studying the scaling laws in wireless networks started by the fundamental work of Gupta and Kumar in \cite{The_Gupta00}, which showed that despite not knowing the exact network capacity from the information-theoretic perspective, we can predict how the per-node throughput scales with the number of nodes under certain physical constraints on the node cooperation and the signal reception.  A more recent approach that is able to better quantify the SINR of a typical network link relies on tools from stochastic geometry \cite{Stochastic_Baccelli10_2} to study the performance of random wireless networks as well-summarized in \cite{Stochastic_ElSawy13}.
 
 
 Particularly relevant to the multi-antenna case are  \cite{Spectral_Lee16,Transmission_Hunter08}, where the authors derived the scaling laws of the transmission capacity in \cite{Transmission_Hunter08} and the ASE in \cite{Spectral_Lee16} for ad hoc networks with multi-antenna nodes, where the number of antennas scales with the density of the nodes. The three key results in \cite{Spectral_Lee16} are: the ASE asymptotically drops to zero in the single-antenna case, super-linear scaling of the number of antennas is required to maintain a linear scaling of the ASE, and nodes cooperation improves the ASE scaling law. Hence, there is a potential to improve the scaling laws of the ASE by increasing the number of antennas.   Interestingly, the scaling laws in cellular networks are different from the ones in \cite{Spectral_Lee16}, as we show in this work.

We assume that the BSs are spatially distributed as a homogeneous Poisson point process (HPPP) with density $\lambda$, the use of any physically feasible path loss model, and independent and identically distributed (i.i.d.) circularly symmetric complex Gaussian channels between the different transmit and receive antennas. For the MISO case with $N_t(\lambda)$ transmit antennas, we prove that the average SINR scales as $\sfrac{N_t(\lambda)}{\lambda}$ and the average ASE scales as $\lambda\log\left(1+\sfrac{N_t(\lambda)}{\lambda}\right)$. Then we generalize the results for the MIMO case with eigen-beamforming, a single stream of data, and $N_r(\lambda)\leq N_t(\lambda)$ receive antennas. We prove that the scaling laws of the conditional SINR and ASE are agnostic to  $N_r(\lambda)$ and scale exactly the same as the MISO case.  Overall, we show that we can maintain a non-zero per-user throughput in dense networks if the number of antennas is properly scaled with the BS density. The full version of this work is available at \cite{Area_AlAmmouri20}. Hence, for the detailed proofs and for more results and discussions, refer to \cite{Area_AlAmmouri20}.

\section{System Model}
{\bf Network Model:} We consider a single-tier downlink cellular network where the BSs are spatially distributed as a two-dimensional HPPP $\Phi$ with intensity $\lambda$ \cite{Stochastic_Baccelli10_2}. Users are spatially distributed according to an independent stationary point process, with intensity $\lambda_u \gg \lambda$, such that each BS has at least one user to serve with probability one. Each BS schedules its users on orthogonal resource blocks such that one user is associated with every BS in a given resource block. Users are assumed to connect to their closest BS, i.e., the BS with the highest average received power. 

The BSs are equipped with $N_t(\lambda)$ antennas, where $N_t$ is a non-decreasing function with $N_t(0)=0$ and $\lim\limits_{\lambda\rightarrow\infty}N_t(\lambda)=\infty$. We consider the three cases where $N_t(\lambda)$ is asymptotically sub-linear, linear, and super-linear, i.e., $\lim\limits_{\lambda \rightarrow \infty}\frac{N_t(\lambda)}{\lambda}=0$, $\lim\limits_{\lambda \rightarrow \infty}\frac{N_t(\lambda)}{\lambda}=c \in \mathbb{R}_{+}$, and $\lim\limits_{\lambda \rightarrow \infty}\frac{N_t(\lambda)}{\lambda}=\infty$, respectively, and for simplicity, we omit the word  {\it asymptotically} when we refer to these cases. Users are equipped with $N_r(\lambda)$ antennas, where we assume that $N_r(\lambda)\leq N_t(\lambda)$, i.e., $\lim\limits_{\lambda \rightarrow \infty} \frac{N_r(\lambda)}{N_t(\lambda)}=y \in[0,1]$.

{\bf Propagation Model:} The large-scale channel gain is assumed to be captured by the function $L:\mathbb{R}_{+}\rightarrow \mathbb{R}_{+}$, i.e., $L^{-1}(\cdot)$ is the path loss. We focus on the class of physically feasible path loss models introduced in \cite{A_AlAmmouri19}, which is characterized by three simple and intuitive properties: ($i$) $L_0:=L(0)$ is a non-zero finite constant, ($ii$) $L(r)\leq L_0, \ \forall r \in[0,\infty)$, and ($iii$) $\gamma:=\int\limits_{0}^{\infty} r L(r) dr$ is also a non-zero finite constant. The first requirement translates to having a finite BS transmit power; the second ensures that the average received power is less than the transmit power, and the third guarantees that the sum of received powers from all BSs is almost surely (a.s.)  finite at any location in the network \cite{A_AlAmmouri19}. We further assume that the path loss function has to satisfy the following assumptions to maintain analytical tractability.

\begin{assumption}
	The path loss function must satisfy the following: $\exists \  r_0 \in {R}_{+}$,  $\zeta \in \mathbb{R}_{+}^{*}$, and a differentiable decreasing function $\tilde{L}: [r_0,\infty)\rightarrow \mathbb{R}_{+}$ such that:
	\begin{enumerate}
		\item $\tilde{L}(r)\leq L(r), \ \forall r \in [r_0,\infty)$.
		\item $\frac{r\tilde{L}(r)}{-\tilde{L}^{'}(r)} \geq \zeta ,\  \forall r \in [r_0,\infty)$.
		\item $\int\limits_{r_0}^{\infty} \frac{r}{\tilde{L}(r)^2} e^{- \pi \lambda_0 r^2}dr$ is finite for all $\lambda_0>\lambda_c\in \mathbb{R_{+}}$.
	\end{enumerate}
\end{assumption}

The bounded single-slope, the bounded multi-slope \cite{Downlink_Zhang15}, and the stretched exponential \cite{SINR_AlAmmouri17} path loss models in addition to the path loss models used in 3GPP standards \cite{3GPP2017} for the entire range of    0.5    to    100   GHz bands, are all included in this class of models and satisfy the three conditions in Assumption 1 \cite{A_AlAmmouri19}. However, the power-law model, i.e., $r^{-\eta}$, is not included in this class due to the singularity at $r=0$.

All small-scale fading variables between any two nodes are assumed to be i.i.d. and independent of the locations of the nodes. We focus on the digital beamforming architecture, where each antenna is connected through a separate RF chain, and hence, the BS has direct access to the channel seen by each antenna. The channel, i.e., the small-scale fading, between any transmit antenna and receive antenna, is assumed to follow i.i.d.  circularly symmetric complex Gaussian random variables with zero mean and unit variance,  which inherently means we assume a rich scattering environment with the proper antenna spacing \cite{Foundations_Heath18}. This assumption is questionable when the network utilizes frequency bands in the mmWave and THz bands since the channels are known to be spatially sparse with a few dominant paths \cite{Millimeter_Rappaport14}. Hence, our model is more suitable for traditional cellular bands.  

\section{Methodology of Analysis}\label{Sec:Meth}
We consider the performance of a user located at the origin. The received signal at the tagged user assuming a serving distance of $r_0$ is
\begin{align}\notag
    y_0&=\sqrt{L(r_0)}\bu^{*}_{0}\bH_{0,0}\bp_{0} s_0 \notag\\
    &+\sum\limits_{r_i\in \Phi \setminus B(0,r_0)} \sqrt{L(r_i)}\bu^{*}_{0}\bH_{i,0}\bp_{i} s_{i}+\bu^{*}_{0}n_0,
\end{align}
where $\bH_{i,j}\in \mathbb{C}^{N_r\times N_t}$ is the channel between the $i^{\rm th}$ BS and the $j^{\rm th}$ user, $\bp_{i}\in \mathbb{C}^{N_t\times1}$ is the precoding (beamforming) vector of the $i^{\rm th}$ BS, $\bu_{i}\in \mathbb{C}^{N_r\times1}$ is the combining vector used by the $i^{\rm th}$ user, $s_i$ is the transmitted symbol from the $i^{\rm th}$ BS, $n_0$ is the zero-mean additive white Gaussian noise with variance $\sigma^2$, and $B(0,r_0)$ is a ball centered at the origin with radius $r_0$. Note that the users are assumed to be ordered such that the $i^{\rm th}$ user is connected to the $i^{\rm th}$ BS. The transmitted symbols from the BSs are assumed to be i.i.d. with zero mean and unit energy. 

By conditioning on the network geometry, the channel gains, and the precoding/combining vectors, the SINR is represented by
\begin{align}\label{Eq:SINR_gen}
   {\rm SINR}(\lambda)&=\frac{L(r_0) |\bu^{*}_{0}\bH_{0,0}\bv_{0}|^2}{\sum\limits_{r_i\in \Phi \setminus B(0,r_0)}L(r_i)|\bu^{*}_{0}\bH_{i,0}\bv_{i}|^2+||\bu_0||_{2}^{2}\sigma^2},
\end{align}
where the dependency on $\lambda$ is captured through the distribution of the serving distance $r_0$\footnote{Note that the serving distance has a probability density function (PDF) of $f_{R}(r_0)=2 \pi \lambda r_0 e^{-\pi \lambda r_0^2}$~\cite{A_Andrews11}.}, the density of interfering BSs, and the number of antennas. Both of the BS and the user are assumed to have perfect knowledge of the channel and design their precoding and combining vectors, respectively, to maximize the SNR at the user. Under the assumption that the elements of $\bH$ are drawn from i.i.d. complex Gaussian random variables with zero mean and unit variance, the BS (user) uses the right (left) singular vector corresponding to the maximum eigenvalue of the matrix $\bH$ as its beamforming (combining) vector, which is referred to as eigen-beamforming.  Based on \cite{Transmission_Hunter08}, the SINR in this case can be expressed as
\begin{align}\label{Eq:SINR_MIMO2}
   {\rm SINR}(\lambda)&=\frac{L(r_0) \phi_{0}^2}{\sum\limits_{r_i\in \Phi \setminus B(0,r_0)}L(r_i)g_i+\sigma^2},
\end{align}
where $\phi_{0}$ is the maximum singular value of the matrix $H_{0,0}$ and $g_i, \forall i \in \{1,2, \cdots\},$ are i.i.d. unit mean exponential random variables independent of $\phi_{0}$. The distribution of $\phi_{0}^2$ has been well-studied and known \cite{Largest_Kang03}, but it does not have a simple form. For the special case of $N_r(\lambda)=1$, i.e., the MISO case, the distribution of $\phi_{0}^2$ reduces to a Gamma distribution with shape $N$ and unit rate, i.e., $\tilde{g}\sim \Gamma(N,1)$ \cite{Downlink_Dhillon13}.

The second performance measure we consider is the  ASE \cite{Area_Alouini99}, which represents the sum throughput for all users per unit area. Given our system model, the conditional ASE is defined as \cite{A_AlAmmouri19}
\begin{align}\label{Eq:ASEGen}
    \mathcal{E}(\lambda) =\lambda \log_2(1+ {\rm SINR}(\lambda)) ,
\end{align}
in bps/Hz/m$^{2}$. Note that the average SINR and the average ASE can be found by averaging \eqref{Eq:SINR_gen} and \eqref{Eq:ASEGen}, respectively, over all channel realizations,  precoding/combining vectors, and network configurations. In terms of scaling laws, Fatou's lemma \cite{Real_Royden88} is very helpful since it shows that $\lim\limits_{\lambda \rightarrow \infty}\mathbb{E}[{\rm SINR}(\lambda)]\geq \mathbb{E}[\lim\limits_{\lambda \rightarrow \infty}{\rm SINR}(\lambda)]$ and $\lim\limits_{\lambda \rightarrow \infty}\mathbb{E}[\mathcal{E}(\lambda)]\geq \mathbb{E}[\lim\limits_{\lambda \rightarrow \infty}\mathcal{E}(\lambda)]$.  Another fundamental lemma that we rely on is given next.
\begin{lemma}\label{Lem:Thm1_1}
Let $L(\cdot)$ be a general physically feasible path loss model, $g_i, \forall i \in \{1,2, \cdots \},$ be a sequence of i.i.d. random variables with unit mean, $\Phi$ be a HPPP with intensity $\lambda$, and $r_n$ be the distance from the origin to the $n^{\rm th}$ closest points in $\Phi$, where $n$ is finite, then
\begin{align}
    \lim\limits_{\lambda \rightarrow \infty} \frac{1}{\lambda} \sum\limits_{r_i \in \Phi \setminus B(0,r_n)}L(r_i) g_i= 2 \pi \gamma \ {\rm a.s.}
\end{align}
where $\gamma=\int\limits_{0}^{\infty} r L(r) {\rm d}r$.
\begin{proof}
Let $\lambda=k \lambda_0$, where $k \in \mathbb{Z}_{+}$ and $\lambda_0 \in \mathbb{R}^{*}_{+}$, and $\tilde{\Phi}$ be a PPP with intensity $k \lambda_0$. Then,
\begin{align}
  &\lim\limits_{k \rightarrow \infty} \frac{1}{k\lambda_0} \sum\limits_{r_i \in \tilde{\Phi} \setminus B(0,r_n)}L(r_i) g_i\notag \\
  &=\lim\limits_{k \rightarrow \infty}\left( \frac{1}{k\lambda_0}\sum\limits_{r_i\in \tilde{\Phi}}g_i L(r_i)-\frac{1}{k\lambda_0}\sum\limits_{j=0}^{n-1}g_j L(r_j)\right)\label{eq:Lem1_1}\\
  &=\lim\limits_{k \rightarrow \infty} \frac{1}{k\lambda_0}\sum\limits_{r_i\in \tilde{\Phi}}g_i L(r_i)\label{eq:Lem1_2}\\
  &=\lim\limits_{k \rightarrow \infty}\frac{1}{k\lambda_0}\sum\limits_{j=1}^{k}\sum\limits_{r_{j,i}\in \Psi_{n}}g_{j,i} L(r_{j,i})\label{eq:Lem1_3}\\
  &=\frac{1}{\lambda_0}\mathbb{E}\left[\sum\limits_{r_{0,i}\in \Psi_{0}}g_{0,i} L(r_{0,i})\right]\label{eq:Lem1_4}\\
  &=2 \pi \int\limits_{0}^{\infty}r L(r)dr=2 \pi \gamma,\label{eq:Lem1_5}
\end{align}
where \eqref{eq:Lem1_1} follows by  adding and subtracting the interference from the $n^{\rm th}$ closest points to the origin, \eqref{eq:Lem1_2} holds since $\frac{\sum\limits_{j=0}^{n-1}g_j L(r_j)}{k\lambda_0}\leq \frac{L_0 \sum\limits_{j=0}^{n-1}g_j }{k\lambda_0}$ which approaches zero a.s. as $k \rightarrow \infty$, given that $n$ is finite, \eqref{eq:Lem1_3} follows using the superposition property of PPPs \cite{Stochastic_Baccelli10_2}, where $\Psi_j, \ \forall j \in \{1,2, \cdots, k\}$ are i.i.d. PPPs with density $\lambda_0$,  \eqref{eq:Lem1_4} follows by using the law of large numbers, and \eqref{eq:Lem1_5} is found by using Campbell's theorem \cite{Stochastic_Baccelli10_2} and the definition of $\gamma$. Finally, since the final result is independent of $\lambda_0$, then we can conclude that $ \lim\limits_{\lambda \rightarrow \infty} \frac{1}{\lambda} \sum\limits_{r_i \in \Phi \setminus B(0,r_o)}L(r_i) g_i=\lim\limits_{k \rightarrow \infty} \frac{1}{k\lambda_0} \sum\limits_{r_i \in \tilde{\Phi} \setminus B(0,r_o)}L(r_i) g_i=2\pi \gamma$, which proves the lemma.
\end{proof}
\end{lemma}

\section{MISO Networks}\label{Sec:MISO}
For this scenario, we focus on the case where we have multi-antenna BSs and single-antenna users. Hence, eigen-beamforming reduces to maximum ratio transmission (MRT)~\cite{Foundations_Heath18} and $\phi_{0}^2$ in \eqref{Eq:SINR_gen} reduces to a Gamma distribution with shape $N$ and unit rate, i.e., $\tilde{g}\sim \Gamma(N,1)$.

\subsection{Scaling Laws}
Before delving into the analysis, it is important to recall that in the single antenna case, the conditional and the mean SINR drop to zero and the conditional and the mean ASE approach a finite constant as $\lambda \rightarrow \infty$ \cite{A_AlAmmouri19}.  For the MISO case,  the asymptotic SINR scaling laws as summarized in the next theorem.
\begin{theorem}\label{Th:SINRDig}
For the MISO case, with $N_t(\lambda)$ transmit antennas, the asymptotic conditional SINR has the following scaling law: $\lim\limits_{\lambda \rightarrow \infty}\frac{\lambda}{N_t(\lambda)}{\rm SINR (\lambda)}=\frac{L_0}{2 \pi \gamma}$ a.s. and the mean SINR is equal to the conditional SINR asymptotically with the same scaling laws.
\begin{proof}
Refer to \cite[Appendix A.]{Area_AlAmmouri20}.
\end{proof}
\end{theorem}

Hence, scaling the number of antennas sub-linearly with the density does not prevent the SINR from dropping to zero for high BS densities. The turning point happens when the number of antennas is scaled linearly with the density. In this case, the SINR approaches a finite constant which is desirable since it guarantees a certain QoS or throughput for the users in the dense regime. This roughly means that we can restore the SINR-invariance  property \cite{A_Andrews11} in dense networks by this scaling. For the ASE, the results are given in Theorem \ref{Th:ASEDig}.

\begin{theorem} \label{Th:ASEDig}
For the MISO case, with $N_t(\lambda)$ transmit antennas, the asymptotic conditional ASE scales as $\lambda \log\left(1+\frac{N_t(\lambda)}{\lambda}\right)$ and the mean ASE, i.e., $\mathbb{E}\left[\mathcal{E}(\lambda)\right]$, has the same scaling law as the conditional ASE.  Specifically:
\begin{itemize}
    \item If $\lim\limits_{\lambda \rightarrow \infty}\frac{\lambda}{N_t(\lambda)}=\infty$, then $\mathcal{E}(\lambda) \rightarrow \infty$ at a scale of $N_t(\lambda)$.
     \item If $\lim\limits_{\lambda \rightarrow \infty}\frac{\lambda}{N_t(\lambda)}=c\in \mathbb{R}^{*}_{+}$, then $\mathcal{E}(\lambda) \rightarrow \infty$ at a scale of $\lambda$.
     \item If $\lim\limits_{\lambda \rightarrow \infty}\frac{\lambda}{N_t(\lambda)}=0$, then $\mathcal{E}(\lambda) \rightarrow \infty$ at a scale  $\lambda\log\left(1+\frac{N_t(\lambda)}{\lambda}\right)$.
\end{itemize}
\begin{proof}
Refer to \cite[Appendix B.]{Area_AlAmmouri20}.
\end{proof}
\end{theorem}
\begin{figure*}[t]
\centering
		\begin{subfigure}{.5\textwidth}				\centerline{\includegraphics[width= 3.1in]{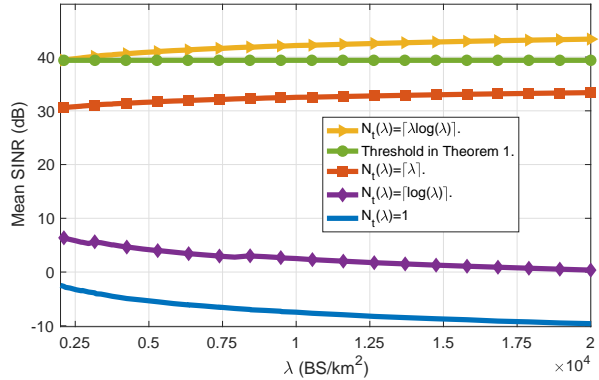}}
		\caption{\, Average SINR.}
	\label{fig:SINR_MISO}
		\end{subfigure}%
			\begin{subfigure}{.5\textwidth}				\centerline{\includegraphics[width= 3.1in]{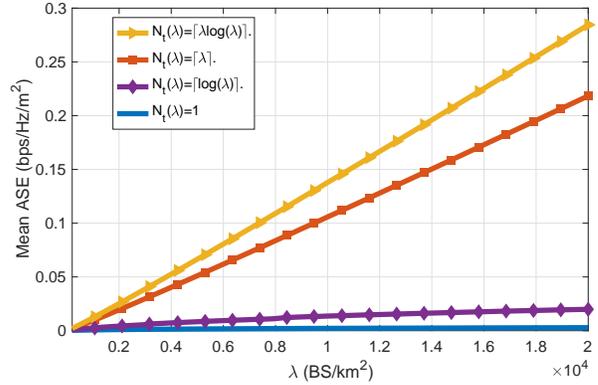}}
		\caption{\, Average ASE.}
		\label{fig:ASE}
		\end{subfigure}
		\caption{Average ASE and SINR vs the BS density $\lambda$ for the MISO case.}
		\label{fig:MISO}
\end{figure*} 

Theorem \ref{Th:ASEDig} shows that although the SINR drops to zero if the number of antennas is scaled sub-linearly with the BS density, we still observe benefits from densifying the network in terms of the sum spatial throughput. This is because the density of the links (users) grows at a rate  faster than the decay of the SINR. Hence, although the throughput of each user tends to zero asymptotically, the sum throughput of all users still grows with densification. Moreover, a linear scaling, which is required to maintain a non-zero SINR, leads to a linear growth of the ASE in dense networks. Overall, the last theorem shows that as long as the number of antennas is scaled positively with the BS density, the densification plateau can be avoided.

\subsection{Numerical Example}\label{Sec:DigSimu}

We start this section by illustrate our derived scaling laws using independent and realistic system level simulations. The simulation uniformly drops BSs in a $20\times20$ km$^2$ region according to the desired density. Then the SINR is evaluated for a user located at the origin. The results were averaged over $10^{4}$ runs. Unless otherwise stated, the noise power is set to $\sigma^2=-70$dBm and the path loss is given by $L(r)=\exp(-\eta r^{-\kappa})$, with $\eta=0.9$ and $\kappa=0.52$. These values were picked since it was shown in \cite{SINR_AlAmmouri17} that using the stretched exponential function accurately captures the path loss in dense urban networks.

Fig. \ref{fig:SINR_MISO} shows the scaling of the mean SINR with the BS density for different scaling rates of the number of antennas; super-linear, linear, sub-linear, and constant (single antenna). We also include the asymptotic value for the linear scaling case given in Theorem \ref{Th:SINRDig}. The curves agree with the derived scaling laws. Precisely, the figure shows that the SINR decreases with the density for the single antenna case, which was proven in \cite{A_AlAmmouri19}, and also when the number of antennas is scaled sub-linearly with the density, which we proved in Theorem \ref{Th:SINRDig}. The figure also shows that a linear scaling of the number of antennas with the BS density is required to prevent the SINR from dropping to zero.


 Moving to the mean ASE, Fig. \ref{fig:ASE}  verifies the scaling laws derived in Theorem \ref{Th:ASEDig} and shows the high gains of densification with antenna scaling in terms of the network throughput. The  figure also highlights the diminishing gains we get by densifying the network when the number of antennas is not scaled with the BS density. These diminishing gains are a result of using a physically feasible path loss model, since for the unbounded power-law model, it was proven that the ASE scales linearly with the BS density \cite{A_Andrews11}. The figure also highlights the linear scaling of the ASE when the number of antennas is scaled linearly with the BS density. 

\begin{figure*}[t]
\centering
		\begin{subfigure}{.5\textwidth}				\centerline{\includegraphics[width= 3.1in]{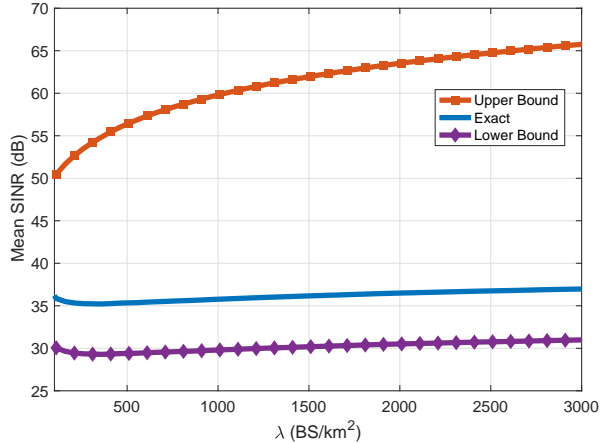}}
		\caption{\, Average SINR.}
		\label{fig:SINR_MIMO}
		\end{subfigure}%
		\begin{subfigure}{.5\textwidth}
        \centerline{\includegraphics[width= 3.1in]{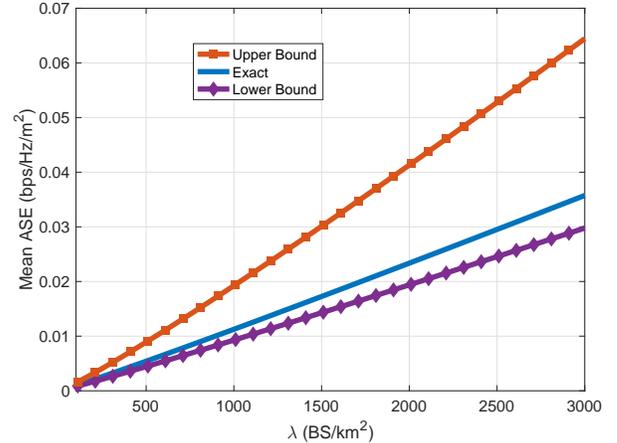}}
		\caption{\, Average ASE.}
		\label{fig:ASE_MIMO}
		\end{subfigure}
		\caption{Average ASE and SINR vs the BS density $\lambda$ assuming eigen-beamforming with $N_t(\lambda)=N_r(\lambda) =\lceil \lambda  \rceil$.}
		\label{fig:MIMO}
\end{figure*} 

\section{MIMO Networks}
Now we go back to the general settings, with $N_r(\lambda)$ receive antennas with $\lim\limits_{\lambda \rightarrow \infty}\frac{N_r(\lambda)}{N_t(\lambda)}=y\in [0,1]$. Recall that we focus on the eigen-beamforming case, where all the antennas are utilized to harness the channel, i.e., only a single stream of data.

\begin{corollary}\label{Cor:MIMO}
For the MIMO case, with $N_t(\lambda)$ transmit antennas and $N_r(\lambda)$ receiver antennas such that $\lim\limits_{\lambda \rightarrow \infty}\frac{N_r(\lambda)}{N_t(\lambda)}=y\in [0,1]$, the conditional SINR has the following scaling law: $\lim\limits_{\lambda \rightarrow \infty}\frac{\lambda}{N_t(\lambda)}{\rm SINR (\lambda)}=\frac{L_0(1+\sqrt{y})^2}{2 \pi \gamma }$ a.s. and the conditional ASE has the same scaling laws as in Theorem \ref{Th:ASEDig} with $N_t(\lambda)$ antennas.
\begin{proof}
Refer to \cite[Appendix C.]{Area_AlAmmouri20}.
\end{proof}
\end{corollary}

Hence, the scaling laws are agnostic to the number of receive antennas and match the scaling laws we derived for the MISO case. More specifically, increasing the number of receive antennas just changes the constant to which $\frac{\lambda {\rm SINR}(\lambda)}{N_t(\lambda)}$ saturates to, but does not change the scaling law. Different from the previous cases, we are unable to derive the exact scaling laws for the average SINR and the average ASE. This is because, to the best of our knowledge, the exact scaling of $\lim\limits_{N_t,N_r\rightarrow \infty}\mathbb{E}[\phi_{0}^2]$ is not known.  Nevertheless, we can derive bounds on the scaling laws as in the following corollary.
\begin{corollary}
For the MIMO case, with $N_t(\lambda)$ transmit antennas and $N_r(\lambda)$ receiver antennas such that $\lim\limits_{\lambda \rightarrow \infty}\frac{N_r(\lambda)}{N_t(\lambda)}=y\in [0,1]$, the average SINR scaling law is faster than $\frac{N_t(\lambda)}{\lambda}$ and slower than $\frac{N_t(\lambda)N_r(\lambda)}{\lambda}$. The average ASE scales at least as the MISO case with $N_t(\lambda)$ antennas and at most as the MISO case with $N_t(\lambda)N_r(\lambda)$ transmit antennas.
\begin{proof}
Refer to \cite[Appendix C.]{Area_AlAmmouri20}.
\end{proof}
\end{corollary}

Hence, for the average SINR, the scaling law is at least similar to the MISO case with $N_t(\lambda)$ antennas and at most similar to the MISO case with $N_t(\lambda)N_r(\lambda)$ antennas. To observe the exact scaling, we use simulations and the results are shown in Fig. \ref{fig:MIMO}, assuming $N_t(\lambda)=N_r(\lambda)=N(\lambda) =\lceil \lambda  \rceil$. Starting by the mean SINR, Fig. \ref{fig:SINR_MIMO} shows that the mean SINR follows the same trend as the lower bound and not the upper bound. In other words, the mean SINR scales as $\frac{N_t(\lambda)}{\lambda}$, which is a constant in this case, and not as $\frac{N_t(\lambda)N_r(\lambda)}{\lambda}$. The results also show that the average ASE scales linearly with the BS density. More specifically, the ASE scaled as $\lambda$, as predicted by the lower bound, and not $\lambda \log_2(\lambda)$, predicted by the upper bound. It also matches the scaling law we derived for and the conditional ASE.
\begin{conjecture}
For the MIMO case with $N_t(\lambda)$ transmit antennas, $N_r(\lambda)$ receive antennas, $\lim\limits_{\lambda \rightarrow \infty}\frac{N_r(\lambda)}{N_t(\lambda)}=y\in [0,1]$, eigenbeamforming, a single data stream, and a physically feasible path loss model that satisfies the requirements in Assumption 1, the average SINR scales as $\frac{N_t(\lambda)}{\lambda}$ and the average ASE scales as $\lambda\log\left(1+\frac{N_t(\lambda)}{\lambda}\right)$.
\end{conjecture}

\section{Conclusion}
In this paper, we have studied the scaling laws in MIMO cellular networks, where the number of antennas scales with the base station (BS) spatial density $\lambda$. In particular, we prove that in the MISO case with $N_t(\lambda)$ transmit antennas,$\sfrac{N_t(\lambda)}{\lambda}$ and the average ASE scales as $\lambda\log\left(1+\sfrac{N_t(\lambda)}{\lambda}\right)$. For the MIMO case with single-stream eigenbeamforming and $N_r(\lambda) \leq N_t(\lambda)$ receive antennas, we prove that the scaling laws of the conditional SINR and ASE are agnostic to $N_r(\lambda)$ and have the same scaling laws as in the MISO case.

\bibliographystyle{IEEEtran}
\bibliography{AhmadRef}
\vfill

\end{document}